\documentclass[aps,pra,preprint,superscriptaddress]{revtex4-1}
\usepackage{graphicx}
\usepackage{float}
\usepackage{amsmath}
\DeclareGraphicsExtensions{.pdf,.jpeg,.png}
\usepackage{array}
\usepackage{mathrsfs}
\usepackage{bm}
\usepackage{color}

\def\##1{\bm {#1}}
\def\=#1{\underline{\underline #1}}
\def\~#1{\tilde{#1}}
\def\^#1{\breve{#1}}
\def\`#1{{#1^\prime}}
\def\:#1{#1^{\prime\prime}}
\def\h#1{\hat{\bm{#1}}}
\def\s#1{\bm{\mathcal{#1}}}

\def\.{\mbox{ \tiny{$^\bullet$} }}

\def\rot{_{\rm rot}}

\begin{document}

%\preprint{}

\title{Lorentz Invariance of Absorption and Extinction Cross Sections of a Uniformly Moving Object}


\author{Timothy J. Garner}
\email[]{tjg236@psu.edu}
%\thanks{}
\affiliation{Department of Electrical Engineering, Pennsylvania State University, University Park, PA 16802}

\author{Akhlesh Lakhtakia}
\affiliation{Department of Engineering Science and Mechanics, Pennsylvania State University, University Park, PA 16802}

\author{James K. Breakall}
\affiliation{Department of Electrical Engineering, Pennsylvania State University, University Park, PA 16802}

\author{Craig F. Bohren}
\affiliation{Department of Meteorology, Pennsylvania State University, University Park, PA 16802}


\date{\today}

\begin{abstract}
The energy absorption  and energy extinction cross sections of an object in uniform translational motion in free space are Lorentz invariant, but the total energy scattering cross section is not. Indeed, the forward-scattering theorem holds true for co-moving observers but not for other inertial observers. If a pulsed plane wave with finite energy density is incident upon an object, the energies scattered, absorbed, and removed from the incident signal by the object are finite.  The difference between the energy extinction cross section and the sum of the total energy scattering and energy absorption cross sections for a non-co-moving inertial observer can be either negative or positive, depending on the object's velocity, shape, size, and composition. Calculations for a uniformly translating, solid, homogeneous sphere show that all three cross sections go to zero as the sphere recedes directly from the source of the incident signal at speeds approaching $c$, whether the material is a plasmonic metal (e.g., silver) or simply a dissipative dielectric material (e.g., silicon carbide).

\end{abstract}

%\keywords{}

\maketitle

\section{Introduction}

The insertion of a stationary object between a distant source of light and a detector leads to a reduction in the energy received in a finite interval of time by the detector. This reduction, often quantified as the extinction cross section, has long been known to be exactly related to a specific component of the amplitude of the forward scattered light when the object is made of a linear material and is illuminated by a plane wave \cite[p.~39]{vdH1957}. An intuitive way to derive this relationship, called the extinction formula \cite[Sec.~4.21]{vdH1957},
was provided by van~de~Hulst in 1949 \cite{vdH1949}. 

By virtue of the principle of conservation of energy,  extinction must be  due to scattering and absorption. 
In 1955, the sum of the total scattering   and  absorption cross sections was related exactly to the co-polarized scattering amplitude in the forward direction by Jones \cite{Jones1955} and Saxon \cite{Saxon1955} independently. Three years later, de~Hoop \cite{deHoop1958} provided a simpler derivation than that of Jones  \cite{Jones1955}  and equivalent to that of Saxon \cite{Saxon1955}. 
This relationship   is variously called the  cross section theorem \cite{deHoop1958},
 the extinction theorem \cite[p.~39]{vdH1957}, the
optical theorem \cite{Newton1982,Jackson1999,Zangwill}, \cite[p.~73]{BH1983}, and the forward-scattering theorem \cite{Karam1982,Beltrami,Baum2007}. We prefer the last of these names.

A plane wave has finite \textit{power} density. If the object is stationary,  power is removed from an incident plane wave by absorption and \textit{elastic} scattering. A plane wave with its amplitude modulated by a pulse has finite \textit{energy} density. We may use the forward-scattering theorem to compute the energy removed from a pulsed plane wave by integrating over the energy removed from each spectral component.  The total energy removed from an incident pulsed plane wave  is equal to the sum of the  energy absorbed by the object and the total energy scattered in all directions.
 
The extinction formula may also be used to compute the energy removed from a pulsed plane wave by a uniformly translating object.  However, scattering is inelastic due to the two-way Doppler shift \cite{Garner2017_JOSAA} except in the forward direction. Accordingly, although the extinction cross section
has to be equal to the sum of the absorption and the total scattering cross sections  in the co-moving inertial frame of reference \cite{Jones1955,Saxon1955,deHoop1958}, that equality is not guaranteed \textit{prima facie} in the laboratory frame of reference which is also inertial. The co-moving frame is attached to the object and therefore translates with the same velocity as the object itself in the laboratory frame.
 
 Therefore, we decided to calculate the total scattering, absorption, and extinction cross sections of a uniformly translating object in free space illuminated by a pulsed plane wave. Since Einstein first investigated electromagnetic scattering by a moving mirror \cite{Einstein1905}\cite{Lorentz1952}, researchers have computed the electromagnetic fields scattered by uniformly translating objects \cite{Yeh1965, Lee1967, Shiozawa1968, Restrick1968, Lang1971, DeZutter1977,Michielsen1981}.  Most of this research has been focused on the scattering of plane waves.    In a recent paper, we implemented a frame-hopping approach to compute backscattered signals from uniformly translating spheres made of 
 silicon carbide (SiC) \cite{Garner2017_JOSAA}, with the incident signal being a pulsed plane wave.  In another paper \cite{Garner2017_AO}, we defined and computed total energy scattering cross sections for uniformly translating arbitrarily shaped objects    in the laboratory frame.  
 
 Adopting the same approach, we calculated 
 \begin{itemize}
 \item[(i)] the total energy scattering, energy absorption, and  energy extinction cross sections of uniformly translating spheres made of silver  and silicon carbide in  the co-moving frame (in which the object is at rest), and
 \item[(ii)] the total energy scattering  and energy extinction cross sections of the same spheres in  the laboratory frame.
\end{itemize}
We found that the energy extinction cross section (as  computed using the extinction formula) is the same in both inertial frames, a conclusion that we confirmed analytically to hold for any object, but the total energy scattering cross section has different values in the two frames. We also found analytically that the energy absorption cross section  must be identical in both frames.

This paper is laid out as follows:  Section II describes the analytical procedures used to compute the three cross sections. Section III presents the  cross sections for homogeneous spheres as  functions of diameter and velocity. The paper concludes with some remarks in Section IV.
 Vectors are indicated by boldface, unit vectors are decorated with carets, and dyadics \cite{Chen} are indicated by double underbars.    Frequency-domain quantities are decorated with tildes, and script symbols are used for time-domain vectors.  

\section{Analytical Procedures}

We used the frame-hopping technique described in detail elsewhere \cite{Garner2017_JOSAA} to compute scattering by a uniformly moving object.  The laboratory frame is denoted by $K'$ with spacetime variables $(\# r', t')$, and the co-moving frame by $K$ with spacetime variables $(\#r, t)$.  The origins and axes of $K'$ and $K$ coincide at time $t' = t = 0$, and the origin of $K$ lies inside the object.  The incident pulse was taken to reach the origin of $K'$ around $t'=0$, The velocity of the object in $K'$ is denoted by $\# v$.  Primed variables are used in $K'$ and unprimed variables in $K$.  

The electric and magnetic fields of the incident pulsed plane wave in $K'$ are given by
\begin{subequations}
\begin{align}
	\s E'_{inc} \left(\bm{r}',t'\right)&= \h E'_{inc} f(\tau')  \\
	\s B'_{inc} \left(\bm{r}',t'\right)&=\frac{\#{\hat{q}}'_{inc}\times \s E'_{inc} \left(\#r',t'\right)}{c},
\end{align}
\end{subequations}
where 
$\tau' = t' - \h q'_{inc}\cdot \#r'/c$,  $\h q'_{inc}$ is the direction of propagation in $K'$, 
$\h E'_{inc}$ is a fixed  unit vector parallel to the electric field, and $c$ is the speed of light in free space. The function
\begin{equation}
f(\tau') =\cos(2\pi \nu_c\tau') F(\tau')
\end{equation}
contains $\nu_c$ as the frequency of the carrier plane wave and
the pulse function $F(\tau')$ has compact support.

If  $f(\tau')$ is square integrable  \cite{CRC}, 
we may calculate the incident signal's energy density $U'_{inc}$ (with units of energy per area) by integrating the magnitude of the instantaneous Poynting vector over all time \cite[p.~66]{Chen}; thus,
\begin{eqnarray}
\nonumber
U'_{inc} &=& \int\limits^{\infty}_{-\infty}\frac{|\s E'_{inc} \left(\# r',t'\right)|^2}{\eta_0} dt' 
\\
&=&\int\limits^{\infty}_{-\infty}\frac{|f(\tau')|^2}{\eta_0} d\tau' < \infty,
\label{U'inc}
\end{eqnarray}
where $\eta_0$ is the  intrinsic impedance of free space.  

The electric and magnetic fields of the incident signal in $K$ are
\begin{subequations}
\begin{align}
	\s E_{inc} \left(\#r,t\right)&= \#E_{inc} f(\psi\tau)   \\
	\s B_{inc} \left(\#r,t\right)&=\frac{\h q_{inc}\times \s E_{inc} \left(\#r,t\right)}{c},
\end{align}
\end{subequations}
where \cite[Eq. (6)]{Garner2017_JOSAA}
\begin{equation}
\left.\begin{array}{l}
\displaystyle{\tau = t - \#{\hat{q}}_{inc}\. \#r/c}
\\[5pt]
\#E_{inc} =\gamma \left[\=L^{-1}\. \h E'_{inc}+\#v\times \left(\frac{\h q'_{inc}\times\h E'_{inc}}{c}\right)\right]
\\[5pt]
\displaystyle{\h q_{inc}=\frac{\h q'_{inc}\.\=L-\gamma{\#v}/c}{\psi}}
\\
\displaystyle{\=L=\=I +(\gamma-1)\h v \h v}
\\
\displaystyle{\h v = \#v/v,\quad v=\vert\#v\vert}
\\
\displaystyle{\psi=\left(1-\frac{\h q'_{inc}\.\#v}{c}\right)\gamma}
\\
\displaystyle{\gamma= \left(1-v^2/c^2\right)^{-1/2}}
\end{array}
\right\}
\end{equation}
and $\=I$ is the identity dyadic.
The magnitude of the incident electric field  in $K$ is  scaled by the Doppler shift $\psi$ relative to that of the incident electric field in $K'$, that is  \cite[Eqs.~(3.60) and (3.61)]{VB1984}, 
\begin{equation}
\label{new-eq5}
|\#E_{inc}|=\psi |\h E'_{inc}|.  
\end{equation}
Hence, the energy density of the incident signal in $K$ is 
\begin{eqnarray}
\nonumber
U_{inc} &=& \int\limits^{\infty}_{-\infty} \frac{|\s E_{inc}(\#r,t)|^2}{\eta_0}dt 
\\
&=&\int\limits^{\infty}_{-\infty} \frac{|\psi \s E'_{inc}(\#r,\psi t)|^2}{\eta_0}dt =  \psi U'_{inc}.
\label{Eq:IncEnrgDens}
\end{eqnarray}
The incident energy density in $K$ is scaled by $\psi$ relative to the incident energy density in $K'$, which matches the ratio of energies  of a single photon in $K$ and $K'$  \cite[p.~31]{Bohm}.  

Let $W'_{sca}$ denote the total scattered energy, $W'_{abs}$ the
absorbed energy, and $W'_{ext}$ the reduction in the  energy received  by a detector in the forward direction in $K'$, when the object is present. 
These energies are finite because of the finite energy density of the incident signal.  Next,
we define the total energy scattering ($C'_{sca}$), energy absorption ($C'_{abs}$), and energy extinction ($C'_{ext}$) cross sections in $K'$ with units of area as 
\begin{equation}
\label{new-eq8}
\left.\begin{array}{l}
C'_{sca} = W'_{sca}/U'_{inc} \\
C'_{abs} = W'_{abs}/U'_{inc} \\
C'_{ext} = W'_{ext}/U'_{inc}
\end{array}\right\}.
\end{equation}
We also define \textit{normalized} total energy scattering ($Q'_{sca}$), energy absorption ($Q'_{abs}$), and energy extinction ($Q'_{ext}$) cross sections as
\begin{equation}
\left.\begin{array}{l}
Q'_{sca} = C'_{sca}/A \\
Q'_{abs} = C'_{abs}/A \\
Q'_{ext} = C'_{ext}/A
\end{array}\right\},
\end{equation}
where $A$ is the projected area of the object (when at rest) on the plane to which
$\h q'_{inc}$ is perpendicular.  
We also define energy cross sections and normalized energy cross sections in $K$ as 
\begin{equation}
\left.\begin{array}{l}
C_{sca} = W_{sca}/U_{inc} \\
C_{abs} = W_{abs}/U_{inc} \\
C_{ext} = W_{ext}/U_{inc}
\end{array}\right\},
\end{equation}
and
\begin{equation}
\left.\begin{array}{l}
Q_{sca} = C_{sca}/A \\
Q_{abs} = C_{abs}/A \\
Q_{ext} = C_{ext}/A 
\end{array}\right\}.
\end{equation}
Standard definitions of cross sections
are based on time-averaged power and time-averaged
power density for monochromatic fields \cite[Sec.~3.4]{BH1983}, whereas our
definitions are based on total energy and energy density.

\subsection{Total Scattered Energy}
In the far zone,  the scattered signal in a specific direction $\h q_{sca}$ in $K$ can be represented as \cite[Eq.~(25)]{Garner2017_JOSAA}
\begin{subequations}
\begin{align}
\nonumber
	\s E_{sca} \left(\h q_{sca} r,t\right)= \frac{\h E_{sca} g(\h q_{sca}; t-r/c)}{r}
	\\ +\, \frac{(\h q_{sca} \times \h E_{sca}) h(\h q_{sca}; t-r/c)}{r} \\
	\s B_{sca} \left(\h q_{sca} r,t\right)=\frac{\h q_{sca}\times \s E_{sca} \left(\h q_{sca} r,t\right)}{c},
\end{align}
\end{subequations}
where the unit vectors $\h E_{sca}$ and $\h q_{sca}$ are mutually orthogonal,
and the functions $g(\h q_{sca}; t-r/c)$ and $h(\h q_{sca}; t-r/c)$
depend on many factors.
 The presence of $\h q_{sca}$  as an argument of these functions  indicates that the scattered signals depend in general on the scattering direction.  The angular energy density of the scattered signal in a given direction $\h q_{sca}$ in units of energy per solid angle is 
\begin{equation}
\begin{gathered}
	\^U_{sca}(\#{\hat{q}}_{sca})=\int\limits^{\infty}_{-\infty}\frac{r^2|\bm{\mathcal{E}}_{sca} \left(\hat{\bm{q}}_{sca} r,t\right)|^2}{\eta_0} dt \\
 = \int\limits^{\infty}_{-\infty}\frac{g^2(\hat{\bm{q}}_{sca}; t-r/c)+h^2(\hat{\bm{q}}_{sca}; t-r/c)}{\eta_0} dt ,
\end{gathered}
\end{equation}
and the total scattered energy in $K$ is 
\begin{equation}
W_{sca} =  \int\limits^{\pi}_{0}\int\limits^{2\pi}_{0}U_{sca}(\theta,\phi)\sin{\theta}d\phi\, d\theta,
\end{equation}
where $\theta$ and $\phi$ are the angles of $\h q_{sca}$.  

The far-zone fields of the scattered signal may be transformed from $K$ to $K'$ to yield \cite[Eqs.~(27) and (29)]{Garner2017_JOSAA}
\begin{subequations}
\begin{align}
\nonumber
	\s E'_{sca} \left(\h q'_{sca} r',t'\right)= \frac{\#E'_{sca} g(\h q_{sca}; (t'-r'/c)\psi')}{r'} \\  
+ \frac{(\h q'_{sca} \times \#E'_{sca}) h(\h q_{sca}; (t'-r'/c)\psi')}{r'} \\
	\s B'_{sca} \left(\h q'_{sca} r',t'\right)=\frac{\h q'_{sca}\times \s E'_{sca} \left(\h q'_{sca} r',t'\right)}{c},
\end{align}
\label{Eq:EstoE's}
\end{subequations}
where \cite[Eq.~(23)]{Garner2017_JOSAA}
\begin{subequations}
\begin{align}
\#E'_{sca} =\frac{\left[\=L^{-1}\.\h E_{sca} -\#v \times \left(\h q_{sca} \times \h E_{sca} \right)/c\right]\gamma}{\h q_{sca} \. \left(\=L\.\h q'_{sca} - \gamma\#v/c\right)}
\label{Eq:E's}
\\
\h q'_{sca}=\frac{\gamma\#v/c + \h q_{sca} \. \=L}{\psi'}
\label{Eq:q's}
\end{align}
\end{subequations}
and
\begin{equation}
\psi'= \left(1+\frac{\h q_{sca}\.\#v}{c}\right)\gamma
\label{Eq:psi'}
\end{equation}
is the Doppler shift from $K$ to $K'$. Note that $\psi'$ depends on
$\h q_{sca}$.

By analogy with Eq.~(\ref{new-eq5}),  the magnitude of the numerator of the
right side of Eq. (\ref{Eq:E's}) is equal to $\psi'$.  The magnitude of the denominator of the
right side of Eq. (\ref{Eq:E's}) may be evaluated by first obtaining the expression for $\h q_{sca}$ in terms of $\h q'_{sca}$ by swapping the primed and unprimed variables in Eq. (\ref{Eq:q's}) 
and replacing $\#v$ by $-\#v$
to obtain 
\begin{equation}
\h q_{sca} = \frac{-\gamma\#v/c+\h q'_{sca}\.\=L}{\left(1-\h q'_{sca}\.\#v/c\right)\gamma}
\end{equation}
and then substituting $\=L\.\h q'_{sca} - \gamma\#v/c=(1-\h q'_{sca}\cdot \#v/c)\gamma\h q_{sca}$; then the magnitude of the denominator of the
right side of Eq. (\ref{Eq:E's}) is $(1-\h q'_{sca} \.\#v/c)\gamma$.  By comparison with  Eq. (\ref{Eq:psi'}), $(1-\h q'_{sca} \.\#v/c)\gamma$ is the Doppler shift from $K'$ to $K$ in the $\h q'_{sca}$ direction and is equal to $1/\psi'$ as the shift from $K$ to $K'$ and back to $K$ in that direction must be unity.  
Since the far-zone scattered field is transverse to the scattering direction in both
$K$ and $K'$,
\begin{equation}
\frac{|\#E'_{sca}|}{|\h E_{sca}|} \equiv \frac{|\h q'_{sca} \times \#E'_{sca}|}{|\h q_{sca} \times \h E_{sca}|} = \psi'^2. 
\end{equation}
The angular energy density of the scattered signal in $K'$ is then 
\begin{equation}
\begin{gathered}
	\^U'_{sca}(\h q'_{sca})= \int\limits^{\infty}_{-\infty} \frac{r'^2|\s E'_{sca}(\h q'_{sca}r',t')|^2}{\eta_0}dt'  \\=\psi'^4
	\int\limits^{\infty}_{-\infty} \left\{ \frac{| g\left[\hat{\bm{q}}_{sca}; (t'-r'/c)\psi'\right]|^2 }{\eta_0} \right. \\
	\left.+ \frac{ |h\left[\hat{\bm{q}}_{sca}; (t'-r'/c)\psi'\right]|^2 }{\eta_0}\right\} dt' ,
\end{gathered}
\end{equation}
which yields 
\begin{equation}
	\^U'_{sca}(\h q'_{sca})	=\psi'^3 \^U_{sca}(\h q_{sca}). 
\label{Eq:U'sca}
\end{equation}
The total scattered energy in $K'$ is
\begin{equation}
	W'_{sca} = \int\limits^{\pi}_{0}\int\limits^{2\pi}_{0}\^U'_{sca}(\theta',\phi')\sin{\theta'}d\phi'\, d\theta', 
\label{Eq:W'sca}
\end{equation}
where $\theta'$ and $\phi'$ are the angles of $\h q'_{sca}$. 

The total scattered energy in $K'$ may now be expressed in terms of the scattered energy density in $K$ by using Eq.~(\ref{Eq:U'sca}). To do this, it is helpful to use two new coordinate systems $K\rot$ and $K'\rot$ that are rotated such that $\#v$ aligns with the unit vectors $\h z\rot$ and $\h z'\rot$, respectively. 
$K\rot$ is co-moving with the object, and its origin coincides with the origin of $K$.  Likewise, $K'\rot$ is stationary with respect to $K'$, and the origins of $K'\rot$ and $K'$ coincide. The origins and axes of $K'\rot$ and $K\rot$ coincide at time $t'=t=0$.  In $K\rot$ and $K'\rot$ respectively, $\psi'$ depends only on $\theta\rot$ and $\theta'\rot$ (measured from the $z\rot$ and $z'\rot$ axes, respectively) and is given in $K\rot$ by
\begin{equation}
	\psi'=(1+\frac{v}{c}\cos\theta\rot)\gamma.
\end{equation}
We define the variables $u\rot = \cos\theta'\rot$ and $w\rot=\cos\theta\rot$ to simplify the integration of the scattered energy density with respect to the directions of $\h q'_{sca}$ and $\h q_{sca}$, respectively. In $K'\rot$,
\begin{equation}
	W'_{sca} = \int\limits^{1}_{-1}\int\limits^{2\pi}_{0}\^U'_{sca}(\cos^{-1}{u\rot},\phi'\rot)d\phi'\rot du\rot.
\end{equation}
Next, we change the angular variables to  $K\rot$  using 
\begin{equation}
\left.\begin{array}{l}
\phi'\rot = \phi\rot
\\[4pt]
\displaystyle{
u\rot=\h q'_{sca} \.\h z\rot=\frac{\cos{\theta\rot}+v/c}{1+(v/c) \cos{\theta\rot}} = \frac{w\rot+v/c}{1+(v/c) w\rot}
}
\\[8pt]
\displaystyle{du\rot = \frac{1-(v/c)^2}{[1+(v/c) w\rot]^2}dw\rot  =dw\rot/\psi'^2}
\end{array}
\right\}. 
\end{equation}
Equation (\ref{Eq:U'sca}) and the foregoing substitutions  transform the right side of Eq.~(\ref{Eq:W'sca}) to yield
\begin{equation}
	W'_{sca}=\int\limits^{1}_{-1}\int\limits^{2\pi}_{0}\psi'(\theta\rot)\^U_{sca}(\cos^{-1}{w\rot},\phi\rot)d\phi\rot \,dw\rot. 
\end{equation}
Finally, we transform  from $K\rot$ to $K$ to get
\begin{equation}
	W'_{sca}=\int\limits^{\pi}_{0}\int\limits^{2\pi}_{0}\psi'(\theta,\phi)\^U_{sca}(\theta,\phi)\sin\theta \,d\phi \,d\theta,
\end{equation}
where we emphasize that the Doppler shift $\psi'$ depends on the direction
of $\h q_{sca}$.

The transformation of the total scattered energy from $K$ to $K'$ depends on the scattering pattern \cite[Chap.~5]{VLV} in $K$ and on the object's velocity.  This contrasts with the transformation of the incident energy density from $K'$ to $K$ in Eq.~(\ref{Eq:IncEnrgDens}), which depends only on the velocity.

The total scattered energy cross sections in $K'$ and $K$ are not equal because 
numerical results in Sec.~\ref{Sec:Results} show that $W_{sca}{\ne}W'_{sca}\psi$. Therefore,
\begin{equation}
C'_{sca} =\frac{W'_{sca}}{U'_{inc}} = \frac{W'_{sca}\psi}{U_{inc}}\neq \frac{W_{sca}}{U_{inc}}=C_{sca}.
\end{equation}
We therefore conclude that \textit{the total  energy scattering cross section is not Lorentz invariant}.

\subsection{Absorbed Energy Computation}\label{Sec:2B}
The object absorbs energy by absorbing a certain number of photons from the pulsed plane wave. The use of photons is convenient, because the same photons will be absorbed in $K'$ as well as $K$.  However, the energy of an absorbed photon in $K'$ differs from the energy of the same absorbed photon in $K$  due to the Doppler shift between the two reference frames.  Suppose that $J$ photons are absorbed during the entire illumination
by the pulsed plane wave.  If the  frequency of the $j$'th photon,
$j\in[1,J]$, is $\nu'_j$ in $K'$, the  frequency of the same photon is $\nu_j=\psi\nu'_j$ in $K$.  The energy of the $j$'th photon is $E'_j = 2\pi\hbar\nu'_j$ in $K'$ \cite[p. 31]{Bohm} and  $E_j = 2\pi\hbar\nu_j = \psi E'_j$ in $K$, where $\hbar$ is the reduced Planck constant.  Thus, the absorbed energy in $K'$ is related to the absorbed energy in $K$ by
\begin{equation}
\label{new-eq28}
	W'_{abs} =    \frac{W_{abs}}{\psi}.
\end{equation}
By virtue of Eqs.~(\ref{Eq:IncEnrgDens}) 
and (\ref{new-eq28}), the total absorbed energy cross sections in $K'$ and $K$ must be equal:
\begin{equation}
	C'_{abs} = \frac{W'_{abs}}{U'_{inc}} = \frac{W_{abs}/\psi}{U_{inc}/\psi} = C_{abs}
	\label{Cabs-invariant}
\end{equation}
We therefore conclude that \textit{the energy absorption cross section is Lorentz invariant}.  

This result differs from the one \cite[Eq.~(61)]{Twersky1971-2} obtained by Twersky, who did not consider that every photon-absorption event must occur in both $K$ and $K'$. Instead, for  eternal planewave illumination, in both frames he determined the time-averaged absorbed power by calculating the
net energy flux across a large sphere surrounding the object. Although this calculation is correct in $K$ \cite{Jones1955,deHoop1958}, no sphere can enclose the object for all time in $K'$.

The energy absorption cross section $C_{abs}$ may be calculated from the standard (i.e., power-based) absorption cross section $\~C_{abs}(\nu_p)$ \cite[Sec.~3.4]{BH1983}
at each angular frequency of the incident signal. Now,
$\~C_{abs}(\nu_p) = \~P_{abs}(\nu_p)/\~U_{inc}(\nu_p)$, where  $\~P_{abs}(\nu_p)$ is the time-averaged absorbed power, and $\~U_{inc}(\nu_p)$ is the irradiance with units of power per area at frequency $\nu_p$.  The incident signal in $K$ is expressed in the frequency domain by taking the discrete Fourier transform \cite{Papoulis} of a discrete-time version of the incident signal with sampling time $t_s$ and duration $Nt_s$ \cite[Eq.~(11)]{Garner2017_JOSAA}; thus,
\begin{equation}
\s E_{inc}(\#r, t) \approx \frac{2}{N}\text{Re}\left[\sum\limits_{p=1}^M \~{\#E}^{inc}_p \exp\left(-i2\pi \nu_p t + \#k^{inc}_p \. \# r\right)\right].
\label{new-eq30}
\end{equation}
The wavevector and wavenumber at each frequency are $\#k^{inc}_p=\h q_{inc} k_p$ and $k_p = 2\pi \nu_p/c$, respectively.  
The integer $M = N/2 - 1$ for even $N$ and $M = (N-1)/2$ for odd $N$.  The sampling rate $1/t_s$ is chosen to be greater than twice the maximum frequency of the incident signal, and the duration $Nt_s$ must be greater than the duration of the scattered signal, which is determined by trial and error for any specific scattering direction.  The irradiance of each frequency component of the incident signal is 
\begin{equation}
\~U_{inc}(\nu_p)=\frac{2|\~E^{inc}_p|^2}{N^2\eta_0}.
\end{equation}

For each frequency $\nu_p$, the absorbed energy can be calculated by multiplying 
$\~C_{abs}(\nu_p)$ by the irradiance $\~U_{inc}$ at that frequency and the duration $Nt_s$ of the incident signal as 
\begin{equation}
\label{Eq:Wabs}
\~W_{abs}(\nu_p) = \~C_{abs}(\nu_p)\,\~U_{inc}\,(\nu_p) Nt_s =  \~C_{abs}(\nu_p)\,\frac{2}{N \eta_0}|\~{\#E}^{inc}_p|^2 t_s.
\end{equation}
The total  energy absorption cross section in $K$ is then computed by summing the absorbed energy over all frequencies of the incident signal and dividing by the incident energy density, i.e.,
\begin{equation}
\label{Eq:Cabs}
C_{abs} = \frac{W_{abs}}{U_{inc}} = \frac{2}{N \eta_0  U_{inc}}\sum\limits_{p=1}^M \~C_{abs}(\nu_p) |\~{\#E}^{inc}_p|^2 t_s.  
\end{equation}
Knowing $C_{abs}$, we can find $C'_{abs}=C_{abs}$.

\subsection{Computation of Energy Removed from Incident Signal}
In $K$ as well as in $K'$, we can compute the energy removed from the incident signal using the procedure to derive the standard (power-based) extinction
cross section \cite[Sec.~4.21]{vdH1957}\cite[]{VB1985}.    
Analogously to Eq.~(\ref{new-eq30}),
the forward-scattered signal in $K'$ can be expressed via the inverse discrete Fourier transform as
\begin{equation}
	\s E'_{sca}(\h q'_{inc}r', t') \approx \frac{2}{N}\text{Re}\left\{\sum\limits_{p=1}^M \frac{\~{\#E}'^{sca}_p(\h q'_{inc})}{r'} \exp\left[-i2\pi \nu'_p t + (2\pi \nu'_p /c)
	 \h q'_{inc} \. \# r'\right]\right\},
\end{equation}
where {$\~{\#E}'^{sca}_p(\h q'_{inc})$} 
is numerically obtained for the $p$-th discrete Fourier component $\~{\# E}'^{inc}_p$ of the incident field in $K'$. Note that $\~{\# E}'^{inc}_p$
corresponds to the $p$-th discrete Fourier component $\~{\# E}^{inc}_p$ of the incident field in $K$ defined in Eq.~(\ref{new-eq30}).

The standard (power-based) extinction cross section in $K'$ is given by \cite[Eq. (8.113)]{VB1985}
\begin{equation}
	\~C'_{ext}(\nu'_p) =  \frac{4\pi}{k'_p}\text{Im}\left\{\frac{\~{\#{E}}'^{sca}_p(\h q'_{inc})\. \left[\~{\# E}'^{inc}_p\right]^\ast}{|\~{\#E}'^{inc}_p|^2} \right\},
	\label{new-eq36}
\end{equation}
where the asterisk denotes the complex conjugate.
The energy extinction cross section 
\begin{equation}
C'_{ext} =  \frac{2}{N \eta_0  U'_{inc}}\sum\limits_{p=1}^M \~C'_{ext}(\nu'_p) |\~{\#E}'^{inc}_p|^2 t_s',
\end{equation}
in $K'$ then emerges from $\~C'_{ext}(\nu'_p)$ by adopting the same way as   for the absorption cross section in Eq.~(\ref{Eq:Cabs}).

In $K$, the standard power-based extinction cross section is \cite[Eq. (8.113)]{VB1985}
\begin{equation}
	\~C_{ext}(\nu_p) =  \frac{4\pi}{k_p}\text{Im}\left\{\frac{\~{\#{E}}^{sca}_p(\h q_{inc})\. \left[\~{\# E}^{inc}_p\right]^\ast}{|\~{\#E}^{inc}_p|^2} \right\}.  
\end{equation}
As  $k_p = \psi k'_p$,  $|\~{\#E}^{inc}_p|=\psi|\~{\#E}'^{inc}_p|$, and
$\~{\#E}^{sca}_p(\h q_{inc}) \. \left[\~{\#E}^{inc}_p\right]^\ast = \psi^3 \~{\#E}'^{sca}_p(\h q'_{inc})\. \left[\~{\#E}'^{inc}_p\right]^\ast$, the identity
\begin{equation}
\~C_{ext}(\nu_p) = \~C'_{ext}(\nu'_p)
\label{new-eq39}
\end{equation} 
follows. The  energy extinction cross section in $K$ is 
\begin{equation}
C_{ext} =  \frac{2}{N \eta_0  U_{inc}}\sum\limits_{p=1}^M \~C_{ext}(\nu_p) |\~{\#E}^{inc}_p|^2 t_s,
\label{new-eq40}
\end{equation}
leading to
\begin{equation}
C_{ext}   = C'_{ext},
	\label{Cext-invariant}
\end{equation}
on using $t_s = t'_{sca} / \psi$, $|\~{\#E}^{inc}_p|=\psi|\~{\#E}'^{inc}_p|$, and $U_{inc} = \psi U'_{inc}$.  We therefore conclude that \textit{the energy extinction cross section is Lorentz invariant}.  

\subsection{Conservation of Energy in Scattering by Moving Objects}
In $K$, the object is stationary and the forward scattering theorem
\cite{Jones1955,Saxon1955,deHoop1958,Karam1982,Baum2007}
\begin{equation}
C_{ext} = C_{sca} + C_{abs}
\label{new-eq41}
\end{equation}
holds;
equivalently, 
\begin{equation}
W_{ext} =W_{sca} + W_{abs}.
\end{equation}
Although $C'_{abs} = C_{abs}$ and $C'_{ext} = C_{ext}$ by virtue of Eqs.~(\ref{Cabs-invariant}) and (\ref{Cext-invariant}), respectively, 
\begin{equation}
C'_{ext} \neq C'_{sca} + C'_{abs}
\end{equation}
follows from Eq.~(\ref{new-eq41})
 because $C'_{sca} \neq C_{sca}$.  
 Therefore the energy removed from the incident signal in $K'$ does not equal the sum of the scattered energy and absorbed energy in $K'$; i.e., 
 \begin{equation}
 W'_{ext} \neq W'_{sca}+W'_{abs}.
 \end{equation}

Scattering in $K'$ is inelastic, because  each spectral component of the scattered signal is altered by a two-way Doppler shift in every scattering direction.  In $K$, no Doppler shift arises and the scattering is elastic.  The inelasticity of  scattering in $K'$ results in a change in the kinetic energy of the object in $K'$.  The  energy removed from the incident signal should be written as 
\begin{equation}
\label{Eq:K'energy}
W'_{ext}=W'_{sca}+W'_{abs}+W'_{mech}, 
\end{equation}
where $W'_{mech}$ is the mechanical work done on the object by the incident signal and has to be equal to the change in the object's kinetic energy.  Thus
$W'_{mech}$  may be obtained using Eq.~(\ref{Eq:K'energy}) after first computing $W'_{ext}$, $W'_{sca}$, and $W'_{abs}$. 
Of course, were we to posit a mechanical work $W_{mech}$ in $K$, we would have $W_{mech}\equiv0$ since the object is immobile in $K$.

\section{Examples: Silver  and silicon-carbide spheres}\label{Sec:Results}
We computed the normalized energy cross sections of uniformly translating spheres made of silver and silicon carbide with diameters (in $K$) ranging from 10 to 500 nm and velocities ranging in magnitude from 0 to 0.9$c$ and directed along either the $\pm \h z'$ or the $\pm \h x'$ directions.  We obtained the frequency-dependent permittivity in $K$ of silver from the measurements of Hagemann \textit{et al.} \cite{Hagemann1975} and of silicon carbide from the measurements of Larruquert \textit{et al.} \cite{Larruquert2011}; see also Ref.~\citenum{RefractiveIndex}. 

The incident signal in $K'$ was taken to be a plane wave with its amplitude modulated by a Gaussian pulse and which travels in the $+z'$ direction with its electric field aligned along the $x'$ axis.  The electric field of the incident signal in $K'$ is 
\begin{equation}
	\s E'_{inc} \left(\#r',t'\right)= \h x' \cos\left(\frac{2\pi c\tau'}{ \lambda'_c} \right) \exp\left(-\frac{\tau'^2 }{ 2\sigma'^2}\right)  ,
\end{equation}
where $\tau' = t'-z'/c$, $\lambda'_c$ is the free-space wavelength of the
carrier plane wave in $K'$, and $\sigma'$ is the width parameter of the Gaussian
pulse.
Equation~(\ref{U'inc}) yields 
the energy density of the incident signal in $K'$ as \cite[Eq.~(3.898.2)]{Zwillinger2015}
\begin{equation}
U'_{inc}
= \frac{\sigma'\sqrt{\pi}}{2\eta_0}\left\{1+\exp\left[- \left(\frac{2\pi c\sigma'}{\lambda'_c}\right)^2 \right] \right\}.
\label{Eq:U'inc}
\end{equation}

In $K$, the incident signal's electric field is 
\begin{equation}
	\s E_{inc} \left(\#r,t\right)= \# E_{inc} \cos\left(\frac{2\pi c}{\lambda_c\tau}  \right)\exp\left(-\frac{\tau^2 }{ 2\sigma^2}\right),  
\end{equation}
where $\lambda_c = \lambda'_c/\psi$ and $\sigma = \sigma'/\psi$.  Thus,
\begin{equation}
U_{inc}
= \frac{\sigma\sqrt{\pi}}{2\eta_0}\left\{1+\exp\left[- \left(\frac{2\pi c\sigma}{\lambda_c}\right)^2 \right] \right\}
\label{Eq:Uinc}
\end{equation}
follows straightforwardly to conform to the identity $U_{inc}=\psi U'_{inc}$.

For every direction $K'$, we computed $\breve{U}'_{sca}$ by rectangular integration of the scattered power density over the duration of the scattered signal.  We used the 41-point Gauss--Kronrod quadrature \cite{Piessens1983}, \cite[pp. 153-155]{Kahaner1989} to integrate $\breve{U}'_{sca}$ over $\theta'$ and 64-point rectangular integration to integrate over $\phi'$ \cite{Kahaner1989}
in order to compute $W'_{sca}$.  The set of nodes for the 41-point Gauss--Kronrod quadrature contains the nodes for the 20-point Gauss--Legendre quadrature as a subset. We checked for convergence of the integration over $\theta'$ using the  error estimate 
\begin{equation}
\delta'_{\theta}=\left|\frac{W'_{sca, GK}-W'_{sca,GL}}{W'_{sca, GK}}\right| ,
\end{equation}
where $W'_{sca, GK}$ is calculated using the 41-point Gauss--Kronrod quadrature and $W'_{sca,GL}$ is calculated using the 20-point Gauss--Legendre quadrature.   For all cases examined, $\delta'_{\theta}<0.21\%$.   
To check for convergence of the integration over $\phi'$, we defined the error estimate  
\begin{equation}
\delta'_{\phi}=\left|\frac{W'_{sca,64}-W'_{sca,32}}{W'_{sca,64}}\right|, 
\end{equation}
where $W'_{sca,32}$ and $W'_{sca,64}$ are computed by 32-point and 64-point rectangular integration, respectively. In all cases we computed, $\delta_{\phi'}<0.03\%$.   

We computed $\~C_{sca}(\nu_p)$ and $\~C_{ext}(\nu_p)$  using analytic expressions emerging from the exact Lorenz--Mie theory  \cite[p.~103]{BH1983}, and we used  the forward-scattering theorem to compute
\begin{equation}
\~C_{abs}(\nu_p) = \~C_{ext}(\nu_p)-\~C_{sca}(\nu_p).
\end{equation}
We computed $\~C'_{ext}(\nu'_p)$ using Eq.~(\ref{new-eq36}) and found it to satisfy the analytically derived
Eq.~(\ref{new-eq39}), thereby numerically confirming the invariance constraint derived as
Eq.~(\ref{Cext-invariant}).
We first determined $C_{abs}$ using Eq.~(\ref{Eq:Cabs}) and then  exploited
 the analytically derived  
Eq.~(\ref{Cabs-invariant}) to determine
$C'_{abs}$. Finally, we calculated $W'_{sca}$ using Eq.~(\ref{Eq:W'sca})
and obtained $C'_{sca}$ using Eqs.~(\ref{new-eq8})$_1$ and (\ref{Eq:U'inc}).

\begin{figure}
\centering
\includegraphics[width=7.0in]{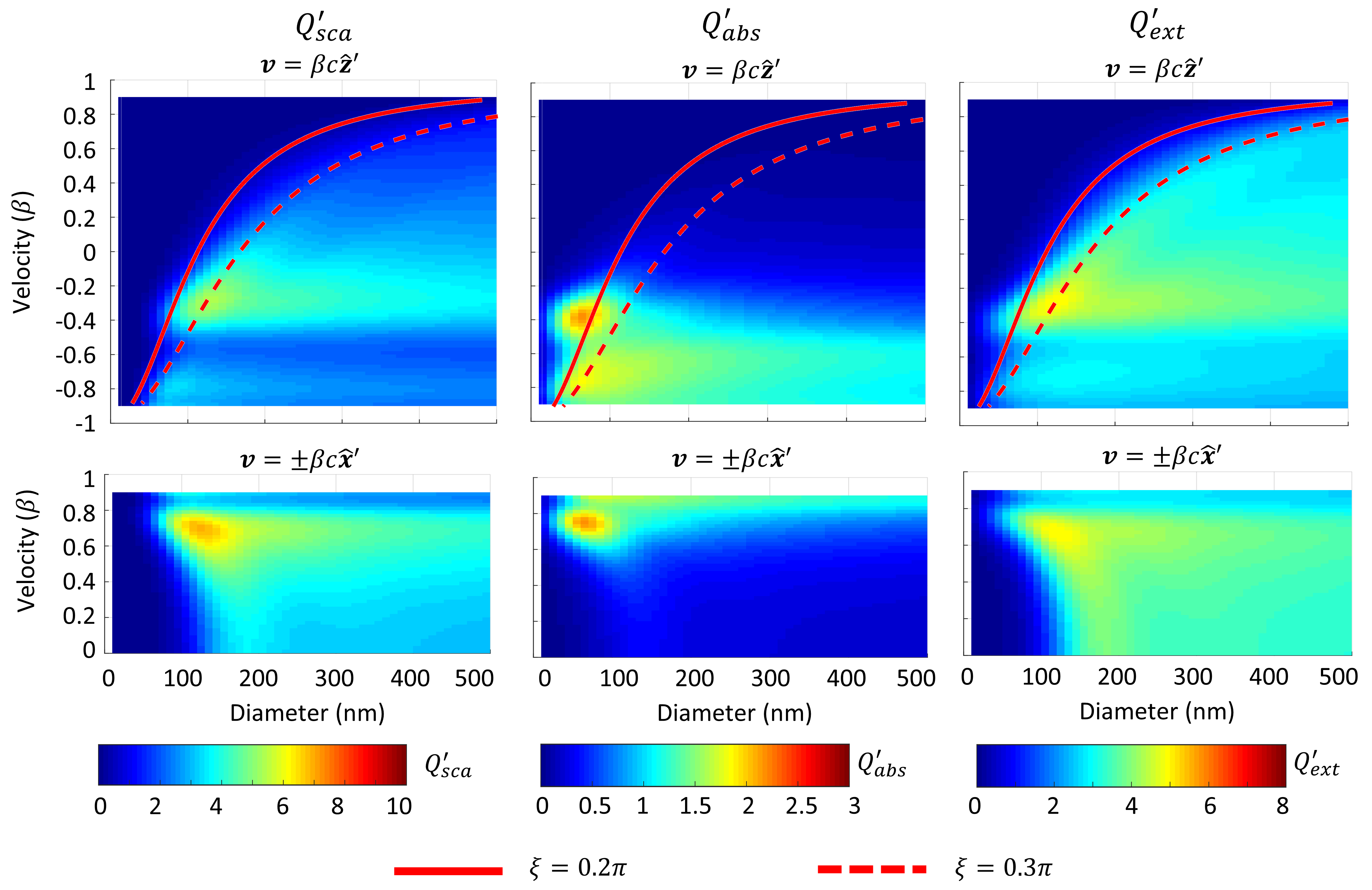}
\caption{(Color online) Normalized total energy scattering ($Q'_{sca}$), energy absorption ($Q'_{abs}$), and eergy extinction ($Q'_{ext}$) cross sections of a silver sphere as functions of $d$ for $\#v\in\left\{\pm\beta{c}{\h x'},  \beta{c}{\h z'}, -\beta{c}{\h z'}\right\}$, $\beta\in[0,0.9]$, when    $\lambda'_c = 550$~nm and $\sigma'=1.83$~fs.}
\label{QAg}
\end{figure}

The normalized total energy scattering, energy absorption, and energy extinction cross sections of a uniformly translating silver sphere are shown in Fig.~\ref{QAg} as functions of diameter $d$ (in $K$) and velocity, where the carrier wavelength of the incident signal in $K'$ is $\lambda'_c = 550 \text{ nm}$, and the width parameter of the Gaussian function is $\sigma'=1.83\text{ fs}$.  The top panels in Fig.~\ref{QAg} show the normalized cross sections when the sphere is directly advancing toward or receding from the source of the incident signal at $\#v = \beta c \h z'$, $\beta \in [-0.9, 0.9]$.  The bottom panels show the normalized cross sections when the sphere is moving transversely to the propagation direction of the incident signal with $\#v = \pm \beta \h x'$, $\beta \in [0, 0.9]$.  A solid and a dashed red line on the top panels show where the electrical size $\xi=\pi d/\lambda_c$ of the sphere in $K$ equals $ 0.2\pi$ and $ 0.3\pi$, respectively.

All three normalized cross sections have a maximum near $d = 100$~nm when $\#v = -0.3c\h z'$.  Both $Q'_{sca}$ and $Q'_{ext}$ rise rapidly as $\xi/\pi$ increases from 0.2 to 0.3.  Due to the symmetry of the sphere, $Q_{abs}$ and $Q_{ext}$ (and thereby $Q'_{abs}$ and $Q'_{ext}$) do not depend on the propagation direction of the incident signal in $K$.  For a fixed $d$, $Q'_{abs}$ and $Q'_{ext}$ depend on velocity only through the Doppler of the incident signal from $K'$ to $K$.  For this reason, the panels that show $Q'_{abs}$ and $Q'_{ext}$ for velocities parallel to  $\pm\h x'$  are affine transformations of a portion of the panels for velocities parallel to  $\pm \h z'$.  This is not true for $Q'_{sca}$, which is demonstrated by the fact that $Q'_{sca}$ takes on greater values in the proximity of $\left\{d = 100\text{ nm},\#v = \pm 0.7c\h x'\right\}$ than for any velocity parallel to  $\pm \h z'$.  

Except for $\#v=\#0$, the normalized energy extinction cross section $Q'_{ext}\ne Q'_{sca}+Q'_{abs}$.  When the sphere recedes directly from the source, $Q'_{ext}>Q'_{abs}+Q'_{sca}$, because the radiation pressure on the sphere increases its kinetic energy.   However, as the radiation force reduces the sphere's kinetic energy when the sphere is advancing toward the source,  $Q'_{ext}<Q'_{sca}+Q'_{abs}$.  Similarly, $Q'_{ext}<Q'_{abs}+Q'_{sca}$ when ${\bf v}\parallel \pm \h x'$, transverse to the propagation direction of the incident signal in $K'$.  The scattering pattern of a moving object gets skewed toward the direction of its velocity.

%%%%%%%%%%%%% Fig. 1 begins %%%%%%%%%%%%%%
%%%%%%%%%%%%% Fig. 1 ends %%%%%%%%%%%%%%

\begin{figure}
\centering
\includegraphics[width=7.0in]{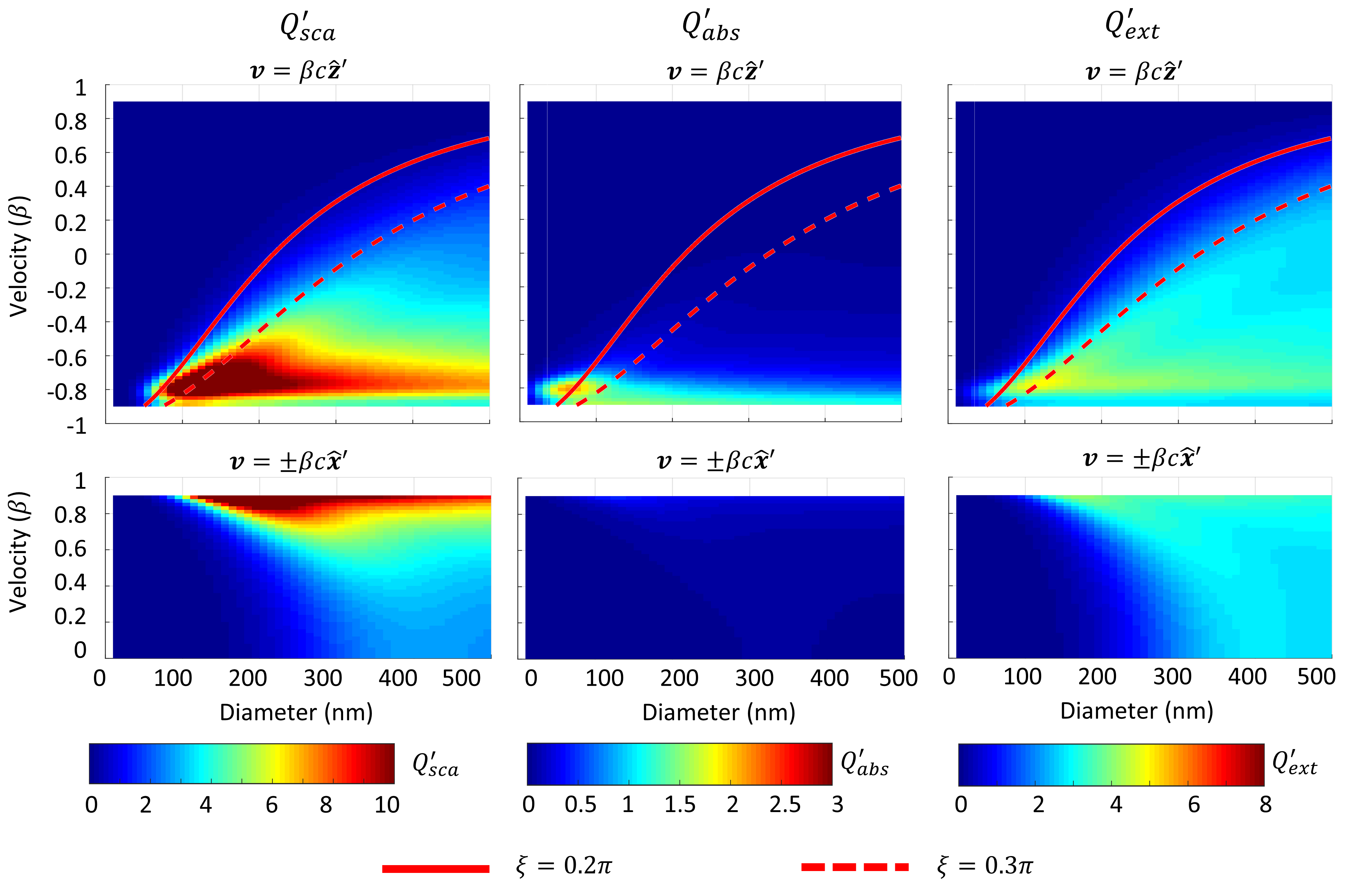}
\caption{(Color online) Same as Fig.~\ref{QAg}, except that
$\lambda'_c = 1100$~nm and $\sigma'=3.67$~fs.}
\label{QAg_1100nm}
\end{figure}

Figure~\ref{QAg_1100nm} shows the normalized total energy scattering, energy absorption, and energy extinction cross sections of a silver sphere when $\lambda'_c = 1100$~nm and $\sigma'=3.67$~fs.  A solid and a dashed line in every panel identify $\xi=0.2\pi$ and $\xi=0.3\pi$, respectively.  As in Fig.~\ref{QAg},
both $Q'_{sca}$ and $Q'_{ext}$   rise rapidly as $\xi/\pi$   increases from $0.2$ to $0.3$.  Each of the three normalized cross sections has a maximum in the
proximity of $\left\{d = 100\text{ nm},\#v=-0.8c\h z'\right\}$.

%%%%%%%%%%%%% Fig. 2 begins %%%%%%%%%%%%%%
%%%%%%%%%%%%% Fig. 2 ends %%%%%%%%%%%%%%

\begin{figure}
\centering
\includegraphics[width=7.0in]{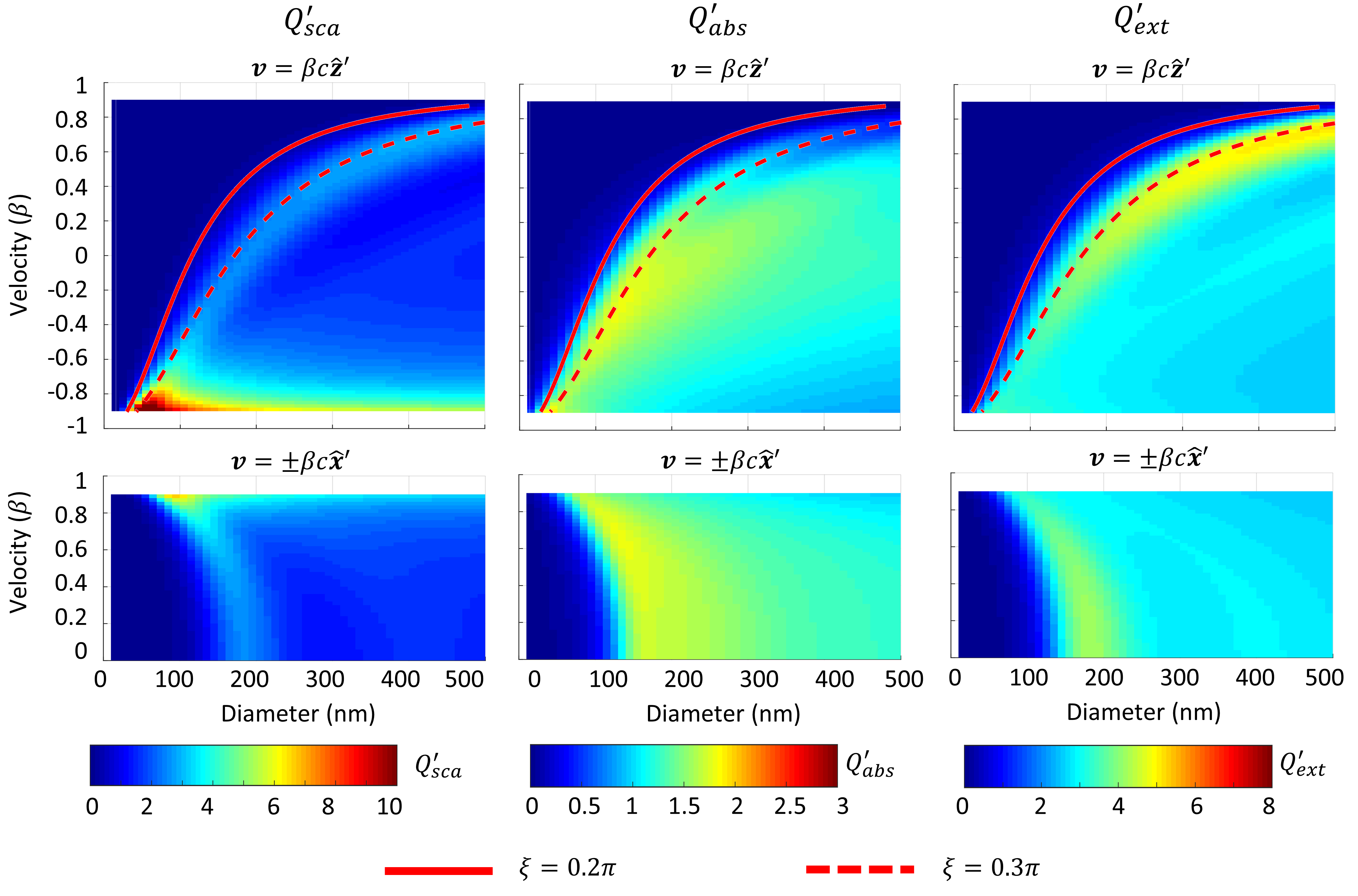}
\caption{(Color online) Same as Fig.~\ref{QAg}, except for a silicon-carbide sphere.}
\label{QSiC}
\end{figure}

 We also computed the normalized cross sections of a silicon-carbide sphere, as shown in Fig. \ref{QSiC}, for an incident signal with $\lambda'_c=550$~nm and $\sigma'=1.83$~fs.  The peak value of $Q'_{sca}$ is found  in the
proximity of $\left\{d = 150\text{ nm},\#v=-0.9c\h z'\right\}$. 
 but the peak value of $Q'_{ext}$  occurs near  $\left\{d = 500\text{ nm},\#v=0.7c\h z'\right\}$.  Thus, as with the silver  sphere, $Q'_{ext}\neq Q'_{sca}+Q'_{abs}$ when $\#v \neq \#0$.  The normalized cross sections rise sharply as $\xi/\pi$   increases from $0.2$ to $0.3$, and both  $Q'_{sca}$ and  $Q'_{ext}$  have a ridge along the $\xi=0.3\pi$ curve.  

%%%%%%%%%%%%% Fig. 3 begins %%%%%%%%%%%%%%
%%%%%%%%%%%%% Fig. 3 ends %%%%%%%%%%%%%%

\section{Concluding Remarks}

We found analytically that the energy absorption  and energy extinction cross sections of an object in uniform translational motion are Lorentz invariant, but the total energy scattering cross section is not.  For that reason, $C'_{ext} \neq C'_{sca} + C'_{abs}$ from the perspective of a stationary observer  (i.e., in $K'$)
but $C_{ext} = C_{sca} + C_{abs}$ from the perspective of a co-moving 
observer (i.e., in $K$).  Thus, the forward-scattering theorem holds true for co-moving observers but not for other inertial observers.

Calculations for a uniformly translating, solid, homogeneous sphere showed
that $C'_{ext}$, $C'_{sca}$, and $C'_{abs}$  depend strongly on the sphere's velocity, size, and composition.  As the sphere recedes directly from the source of the incident signal at speeds approaching $c$, its electrical size in $K$ goes to zero, which   causes the normalized cross sections to go to zero.  Whether the material is a plasmonic metal (e.g., silver) or simply a dissipative dielectric material (e.g., silicon carbide), the normalized cross sections are small when the sphere is electrically small in $K$; furthermore, the normalized cross sections  increase rapidly as the electrical size increases from $0.2\pi$ to $0.3\pi$.  

In terms of energy,  $W'_{ext} \neq W'_{sca} + W'_{abs}$ whereas $W_{ext} = W_{sca} + W_{abs}$.
 The inequality $W'_{ext} \neq W'_{sca} + W'_{abs}$ is due to the increase or decrease in the energy of the scattered photons relative to the incident photons, depending on the two-way Doppler shift in each scattering direction.  The difference between $W'_{ext}$ and $W'_{sca} + W'_{abs}$
  may be accounted for by the change in the object's kinetic energy $ W'_{mech}$. We found numerically that the change in kinetic energy is positive for directly receding spheres, but it is negative for directly advancing spheres and spheres moving transversely to the propagation direction of the incident signal.  

\section*{Acknowledgments}
T. J. G. was supported by a Graduate Excellence Fellowship from the College of Engineering, Pennsylvania State University.  A. L. is grateful for the support of the Charles Godfrey Binder Endowment at the Pennsylvania State University.

\end{document}